\documentclass[10pt, a4paper]{article}

\usepackage{helvet}

\usepackage{lrec-coling2024} 

\usepackage{placeins}

\usepackage{stackengine}
\newcommand\xrowht[2][0]{\addstackgap[.5\dimexpr#2\relax]{\vphantom{#1}}}

\usepackage{natbib}
\usepackage{multibib}
\makeatletter
\def\@mb@citenamelist{cite,citep,citet,citealp,citealt,citepalias,citetalias}
\makeatother
\newcites{languageresource}{~}

\usepackage{graphicx}
\usepackage{tabularx}
\usepackage{soul}
\usepackage{titlesec}
\titleformat{\section}{\normalfont\large\bfseries\center}{\thesection.}{1em}{}
\titleformat{\subsection}{\normalfont\SmallTitleFont\bfseries\raggedright}{\thesubsection.}{1em}{}
\titleformat{\subsubsection}{\normalfont\normalsize\bfseries\raggedright}{\thesubsubsection.}{1em}{}
\renewcommand\thesection{\arabic{section}}
\renewcommand\thesubsection{\thesection.\arabic{subsection}}
\renewcommand\thesubsubsection{\thesubsection.\arabic{subsubsection}}

\usepackage{xcolor}
\usepackage{hyperref}
 \definecolor{darkblue}{rgb}{0, 0, 0.5}
  \hypersetup{colorlinks=true, citecolor=darkblue, linkcolor=darkblue, urlcolor=darkblue}

\usepackage{xstring}

\usepackage{color}

\newcommand{\anonymize}[1]{#1}

\newcolumntype{A}{>{\centering}p{0.03\textwidth}}

\title{InaGVAD : a Challenging French TV and Radio Corpus Annotated for Speech Activity Detection and Speaker Gender Segmentation}

\name{\anonymize{David Doukhan$^{\ast}$, Christine Maertens$^{\ast}$, William Le Personnic$^{\ast}$,}\\ {\bf \large \anonymize{Ludovic Speroni$^{\ast}$, Reda Dehak$^{\dagger}$}}} 

\address{\anonymize{$^{\ast}$French National Institute of Audiovisual (INA), Paris, France} \\
        \anonymize{$^{\dagger}$ Laboratoire de Recherche de l’EPITA (LRE), Paris, France }\\
         \anonymize{\{ddoukhan, cmaertens, wlepersonnic, lusperoni\}@ina.fr, reda.dehak@epita.fr}\\
         }

\abstract{ 
InaGVAD is an audio corpus collected from 10 French radio and 18 TV channels categorized into 4 groups: generalist radio, music radio, news TV, and generalist TV.
It contains 277 1-minute-long annotated recordings aimed at representing the acoustic diversity of French audiovisual programs and was primarily designed to build systems able to monitor men's and women's speaking time in media.
inaGVAD is provided with Voice Activity Detection (VAD) and Speaker Gender Segmentation (SGS) annotations extended with overlap, speaker traits (gender, age, voice quality), and 10 non-speech event categories.
Annotation distributions are detailed for each channel category.
This dataset is partitioned into a 1h development and a 3h37 test subset, allowing fair and reproducible system evaluation.
A benchmark of 6 freely available VAD software is presented, showing diverse abilities based on channel and non-speech event categories.
Two existing SGS systems are evaluated on the corpus and compared against a baseline X-vector transfer learning strategy, trained on the development subset.
Results demonstrate that our proposal, trained on a single - but diverse - hour of data, achieved competitive SGS results.
The entire inaGVAD package; including corpus, annotations, evaluation scripts, and baseline training code; is made freely accessible, fostering future advancement in the domain.
\\ \newline  }

\begin{document}

\maketitleabstract

\section{Introduction}


Voice Activity Detection (\textbf{VAD}) is a signal processing method in charge of identifying segments of audio recordings containing human speech, distinguishing them from portions of the signal containing silence, respiration, background noise, or music.
This method has a wide range of applications in automatic speech analysis: Automatic Speech Recognition (ASR), speaker recognition, speaker traits analysis, and mobile communication~\cite{sharma2021recent, sharma2022comprehensive}.
While recent end-to-end ASR do not depend on isolated VAD anymore~\cite{radford2023robust}, VAD is still useful for detecting generative hallucinations, obtaining precise time-coding of speech transcripts, localizing pauses, disfluencies and speakers-related information.

Speaker gender segmentation (\textbf{SGS}) consists of distinguishing parts of speech segments containing male speech from those containing female speech. Early SGS systems were used in ASR for selecting gender dependent acoustic models allowing to improve Word Error Recognition rates~\cite{lamel1995phone}.
Over the past few years, a growing amount of digital humanity studies, as well as French audiovisual regulation authorities reports, have used automatic SGS  for estimating women's and men's speech percentages in massive amounts of audiovisual media.
With respect to the high social impact and mediatic coverage offered to these studies, SGS systems were used to analyze controlled conditions materials in order to limit gender speaking time estimation errors~\cite{geena, doukhan2018describing,gay2023radio, arcom23}.
While \cite{doukhan2018open} reported SGS evaluation obtained on REPERE~\cite{giraudel2012repere} for the estimation of men and women speaking time percentage, less is known on SGS robustness used on the diversity of audiovisual materials.

The present contribution aims at providing a freely-available annotated speech resource representing the diversity of French audiovisual broadcasts,  
allowing to design and evaluate VAD and SGS systems, and contributing to a better understanding of audiovisual phenomena challenging automatic systems.

This paper is organized as follow. Section \ref{sec:related} presents existing speech resources relevant for building VAD or SGS systems. Section \ref{sec:audiovisualcorpus} presents inaGVAD audiovisual materials and annotation process. Section \ref{sec:corpusanalysis} details inaGVAD annotation distributions. Section \ref{sec:eval} presents a benchmark of 6 open-source VAD and 3 SGS systems on inaGVAD, including an original transfer learning approach. Lastly, section \ref{sec:ccl} provides a discussion on the main contributions, findings, and limitations of our proposal.
Appendices \ref{sec:avail} and \ref{sec:ethical}  finally detail the procedure for downloading inaGVAD 
together with the ethical considerations related to the constitution and diffusion of this corpus.









\section{Related Work}
\label{sec:related}

Speech corpora designed for Automatic Speech Recognition (ASR) generally provide speaker turn and gender annotations, but tend to favor the quantity of annotated lexical terms to the accurate timing of non-speech events (shortest pauses, respiration, sneezing ...), limiting their relevance for building and evaluating VAD and SGS systems.
ESTER 2~\citeplanguageresource{galliano2009ester} and REPERE~\citeplanguageresource{giraudel2012repere} are widely adopted audiovisual speech  resources in the French language for building ASR. 
Their programs 
are mostly composed of news or debates, excluding documentaries, movies and cartoons. 
Among annotated recordings, segments associated to advertisements, inter-program or musical interludes are ignored using UEM\footnote{NIST Unpartitioned Evaluation Map (UEM) format define audio regions that systems must process.} annotations, resulting in materials hardly suited to the analysis of speech detection False Alarms.



Speech resources suited to VAD do provide more accurate timings, but generally lacks speaker traits (gender, age), speech quality (timbre, elocution) and non-speech event (noise, music, respiration) annotations, limiting the description of the nature of phenomena challenging VAD systems.
AVA-Speech corpus~\citeplanguageresource{Chaudhuri2018, gu2018ava} was obtained from Youtube movies (excluding animated) and annotated using a set of 3 labels for describing speech segments (clean speech, speech + noise, speech + music) but using a single label for annotating non-speech events. 
DIHARD 2 \citeplanguageresource{ryant2019second} was realized in the context of challenges focusing on "hard" diarization (meetings, outdoor recordings...) and include a small portion of audiovisual material limited to broadcast news. 
RATS corpus~\citeplanguageresource{rats} contain conversational telephone speech pre-annotated using automatic VAD procedures 
and manually corrected in a 2 pass process.
LibriParty 
 is a synthetic VAD dataset generated with a SpeechBrain recipe~\cite{speechbrain}, based on   LibriSpeech~\citeplanguageresource{panayotov2015librispeech}  read English speech excerpts
 artificially mixed with room impulse responses and environmental noises.


While speaker recognition corpora can be considered for training SGS, they provide isolated speaker segments, not allowing to evaluate speaker changes, nor system behavior in case of speech False alarms. 
Moreover, largest audiovisual-like speaker corpora were built using semi or fully automatic procedures suited to materials sharing several common characteristics (most commonly interviews), using automatic preprocessings (VAD, diarization) discarding atypical vocal performances or noise conditions \citeplanguageresource{salmon2014effortless, voxceleb1, voxceleb2, uro2022semi}.\\



Our proposal is aimed at closing the gap between ASR, VAD, and speaker corpora and provides:
\begin{itemize}
\item fine-grained time-coded speech and non-speech events annotations
\item speaker traits (gender, age) and speech quality annotations
\item materials representing the diversity of contents that can be found in French TV and radio
\item freely available corpus and evaluation code allowing to train and evaluate models in same conditions than in this study
\end{itemize}

\section{Audiovisual Corpus}
\label{sec:audiovisualcorpus}

\subsection{TV and Radio Data}

This audiovisual data corpus was obtained with the support of French National Audiovisual Institute\footnote{\url{https://www.ina.fr}} (INA).
INA is a French public institution in charge of the digitization, preservation, distribution and dissemination of the French audiovisual heritage.
The corpus is composed of one-minute French TV or radio excerpts randomly selected from 24/7 broadcast between January~1, 2021, and December 31, 2022 on 10 radio and 18 TV channels. 
The extracted excerpts are intended to encompass a substantial range of diversity within broadcast materials, including categories that are typically underrepresented in annotated speech corpora: cartoons, fictional content, sports programs, reality TV, documentaries, music, games, inter-program (teasers, advertisements, jingles) together with more documented materials such as talk shows or news.
Resulting audio excerpts were transcoded to mono-channel 16kHz WAV files. By the end of this study, 277 annotated audio files were selected for a public release.

\subsection{Repetition Analysis}

A repetition analysis procedure was carried on to avoid finding identical audio excerpts in the corpus (advertisements, jingles, reruns) using INA's audio fingerprint technology~\cite{chenot2014large}. This analysis, allowing the detection of repetitions associated with a minimal duration of 5 seconds, revealed a solitary repetition in the 277 audio files: a brief 8-second radio jingle broadcast on the same channel between two excerpts. 
These two duplicated excerpts are in the testing set (see section \ref{subsec:devtest}), ensuring the same signal portions are not duplicated in the development and in the testing set.

\subsection{Channel Categories}
TV and radio channels were categorized into 4 groups: generalist TV, news TV, generalist radio, and music radio.

The \textbf{news TV} group refers to 4 continuous TV news channels, a material widely represented in annotated speech resource associated with high amounts of speech overlaps and degraded recording conditions (phone, etc..): BFM TV, CNews, France 24 (public with worldwide broadcast area), and LCI.

The \textbf{generalist TV} category includes 14 channels having a generalist or thematic status: Arte (Franco-German), Canal+ (broadcasting a large amount of fictional movies and sports content), Chérie 25 (thematic women-oriented channel), France 2, France 3, France 4 (child-oriented programs on day time and music programs at night), France 5 (documentaries and educative programs), Gulli (child-oriented programs), M6, NRJ 12 (teenage-oriented TV-reality programs and series), Paris Première, TV5 monde (international channel in French language with a worldwide broadcast area), TF1 (French generalist TV having the largest audience), and TFX (series). The choice of grouping thematic and generalist channels within the same category was motivated by the difficulty of using channel status denomination for describing their actual content and acoustic properties: France 4 has a generalist status despite its large amount of child-oriented and musical programs, whereas Chérie 25 has a thematic status together with a program grid similar to several generalist channels.


Two groups of radio channels were defined based on the estimated amount of broadcast music reported in~\cite{doukhan2018describing}.
\textbf{music radio} category is used for 6 channels having more  than 50\% of musical content: France Bleu (generalist), FIP (eclectic), France Musique (classical), Fun radio (teenage oriented, pop), Mouv' (hiphop), Skyrock (pop, hiphop). Lastly, the \textbf{generalist radio} category is used for the 4 remaining radio channels with more than 50\% of speech: France Culture, France Info, RMC (large amount of sports programs), and RTL.

These four categories present different challenges for automatic VAD and SGS. We expect more speech overlapping in news programs linked to debates; music radio and generalist TV categories will present different backgrounds and noise.



\subsection{Annotation Scheme}
\label{sec:annotscheme}

A speech and non-speech annotation scheme was defined to help the description of phenomena challenging speech activity detection and gender prediction systems.
In order to have a reasonable trade-off between annotation quality and quantity, annotators were instructed to segment audio streams using segments with a minimum duration of 300 milliseconds.

\subsubsection{Speech Segment Coding}

Speech segments were defined as portions of an audio signal containing audible and intelligible spoken voice.
This definition excludes singing voice, hubbubs and non intelligible low-volume speech.
Speech segments were annotated using a sequence of 2 to 3 characters detailed in table~\ref{tab:speech} related to speaker gender, speaker age, and optionally, speech quality.

Speaker gender coding was performed using three categories: male, female, and IDK (I don't know).
As long as naive listeners have strong tendencies to assign voices to binary gender categories~\cite{doukhan2023voice}, annotators were strongly encouraged to use IDK labels when facing hesitations.
They were also instructed that children and elderly character's voices found in cartoons and dubbed programs are often dubbed by adult actors, not necessarily having  the same gender nor the same age as the character being dubbed~\cite{doukhan2018describing} and encouraged to use IDK gender label when facing such vocal performances.
Lastly, annotators were told that IDK gender annotations could be highly valuable for improving automatic gender prediction technologies and quantifying the limits of speech-based gender monitoring in media. 

Speaker age  was coded using three categories: Child (defined as prepubescent voice), adult, and elderly: gender perception and automatic prediction from voice being known to be more challenging for children and elderly persons voices~\cite{schuller2013paralinguistics}.

The last optional information on voice quality distinguishes interjections or onomatopoeia (ah, oh, eh, aie) from regular lexical content. It also allows to distinguish speech segments with intelligible lexical content uttered with atypical voice quality: crying, laughing or shouted speech, ill person voice (cold, tracheotomy), artificially distorted voices (auto-tune, anonymity voices), vocal performance (cartoon character or monster voice).

When facing speech segments uttered by distinct speakers sharing the same speaker coding (\textit{i.e.} 2 adult females without pause between speech turns), annotators were asked to use a single speech segment in order to lower total annotation time.

\begin{table}
\centering
\begin{tabularx}{0.5\textwidth}{|l|X|}
\hline  
\multicolumn{2}{|l|}{\xrowht[()]{10pt} \textit{Speaker gender coding (first character)}} \\ \hline
\textbf{H} & Male (Homme)\\ \hline 
\textbf{F} & Female (Femme) \\ \hline
\textbf{I} & I Don't Know / IDK (Inconnu) \\ \hline
\multicolumn{2}{|l|}{\xrowht[()]{10pt} \textit{Speaker age coding (second character)}}\\ \hline
\textbf{A} & Adult\\ \hline
\textbf{E} & Child / Prepubescent   (Enfant) \\ \hline
\textbf{S} & Ederly (Senior) \\ \hline
\multicolumn{2}{|l|}{\xrowht[()]{10pt} \textit{Speech quality coding (3rd character - optional)}}\\ \hline
\textbf{\{\}} & empty : standard \\ \hline
\textbf{-} & Interjection or onomatopoeia \\ \hline
\textbf{*} &  Atypical voice quality\\ 
\hline
\end{tabularx}
\caption{\label{tab:speech} Speech Segment coding scheme. Symbols acronyms are inferred from French words indicated within parentheses}
\end{table}

\subsubsection{Non-Speech Segment Coding}

\begin{table}
\centering
\begin{tabular}{ll}
\hline
\textbf{Symb} & \textbf{Description} \\
\hline
\textbf{AP} & Applause and claps (APplaudissements)\\
\textbf{BR} & environmental noise (BRuit)\\
\textbf{BH} & hubbub (BrouHaha)\\
\textbf{JI} & JIngle \\
\textbf{MU1} & foreground MUsic\\
\textbf{MU2} & background MUsic \\
\textbf{RE} &  REspiration\\
\textbf{RI} & non-intelligible laughers (RIres) \\
\textbf{AU} & other (AUtre)\\ 
\textbf{\{\} }& empty\\
\hline
\end{tabular}
\caption{\label{tab:nonspeech} Symbols used for the annotation of non-speech events. Symbols acronyms are inferred from French words indicated within parentheses}
\end{table}

Non-speech events were annotated using a set of 10 labels listed in table \ref{tab:nonspeech}. This coding scheme includes 3 categories of musical events: foreground music (including singing voice without instruments), background music (mostly overlapped with spoken speech and annotated between breath groups) and jingles. Three additional non-speech event categories refer to sounds produced by human beings: respiration (including inspirations and  audible sniffing), laughs, and hubbub (defined as non intelligible speech sounds uttered by a crowd or a group of individuals). Noise events were described using three categories: applause, environmental noise (falling objects, alarm, motors, animal sounds), and other (audible non-speech events that can be assigned to any other categories). Lastly, non-speech segments containing no audible or distinguishable sounds were left empty.

\subsubsection{Overlapping Phenomena Coding}

Annotators were instructed to represent overlapping phenomena using \texttt{'+'} symbol (\textit{e.g.} \texttt{'HA*+FE'} being used for overlapping speech uttered simultaneously by an adult male with atypical voice quality and a female child; \texttt{'MU2+RE'} corresponding to audible respiration mixed with background music).
Non-speech events occurring concurrently with speech events were not annotated; for instance, adult female speech mixed with noise is only annotated with a \texttt{'FA'} symbol.
This annotation instruction is aimed to lower the cost and complexity of the annotation campaign.
It is also motivated by a greater need for resources allowing  to describe VAD False Alarm errors based on non-speech event categories, rather than describing gender prediction errors based on overlapping non-speech event categories.

\subsection{Annotation Procedure}
\label{sec:annotationprocedure}

\begin{figure*}
\begin{center}
\includegraphics[width=\textwidth]{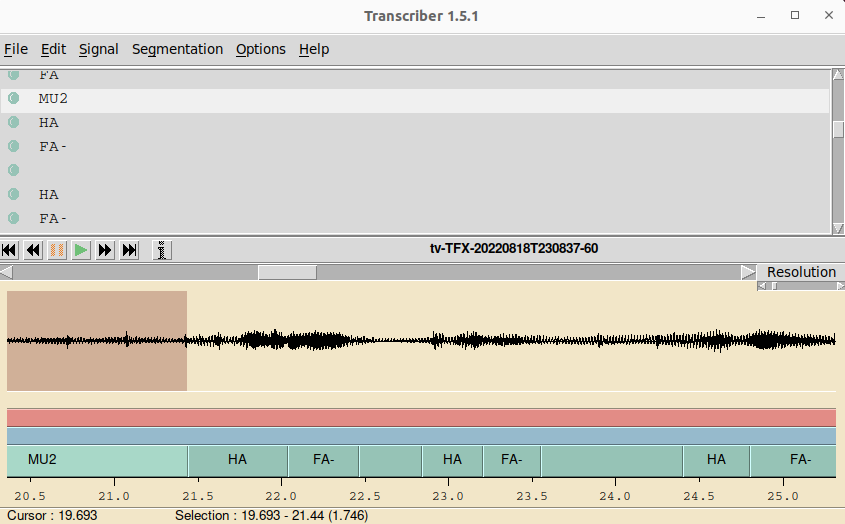} 
\caption{Speech and non-speech event annotation in Transcriber based on the following labels : MU2 (background music), HA (adult male spoken voice), FA- (adult female uttering onomatopoeia) and empty label (nothing relevant)}
\label{fig:transcriber}
\end{center}
\end{figure*}

The corpus annotation task was carried out 
by a 56-year-old female audiovisual archivist and a 28-year-old male multimedia data technician.
TV programs video tracks were not provided to annotators, to ensure their judgments would only be based on auditory cues and not be biased by visual elements (face aspect, etc...).

Annotators were initially briefed on fundamental concepts in audio machine learning such as segmentation, ASR, diarization along with the utilization of audio annotation softwares.
They were also introduced to collar-based segmentation evaluation metrics~\cite{pyannote.metrics} and instructed that resulting annotations will be used using collar values of 300 ms (ignoring zones 150ms before and  after segment boundaries). They were also encouraged to refrain from annotating segments of duration lesser than 300 ms that would be disregarded during evaluations and would increase annotation time.

Each annotator devoted 48 hours (6 workdays) to the project, encompassing training, meetings, and annotation, totaling 96 working hours. 
Annotators were asked to prioritize annotation quality over quantity and to work under ergonomic conditions, using a 
high-quality audio headset, a large keyboard, a mouse and a spacious screen (no laptop).
Annotations were realized using Transcriber open-source software~\cite{barras1998transcriber}, allowing the display of sound file wave forms aligned with annotations. As illustrated by figure~\ref{fig:transcriber}, 
annotators entered speech and non speech event labels in annotation layers usually dedicated to speech transcripts, which were found to be the most ergonomic option to achieve their goals.
They were asked to upload their contributions (transcriber \texttt{.trs} files) to an external Flask-based website\footnote{\url{https://flask.palletsprojects.com}} realized to help improving annotation quality and reliability.
Upon upload, segments with labels containing syntax errors were listed (see annotation scheme in sec \ref{sec:annotscheme}) and 
rejected awaiting annotator corrections.
Warnings related to small segment duration and adjacent segments sharing the same labels were also issued, as these phenomena were often 
associated with annotation errors.
In addition, 
visual difference between manual and automatic \texttt{inaSpeechSegmenter}~\cite{doukhan2018open} annotations were displayed 
together with VAD and SGS evaluation metrics detailed in section \ref{sec:eval}.
Annotators were asked to inspect significant differences between manual and automatic analyses, and correct in Transcriber 
obvious annotation errors.
Conversely, they were asked to refrain from being influenced by machine judgments in case of personal perception differences. 



By the end of this process, 285 minutes of annotated audio content were produced within an involvement time of 96h, corresponding to a ratio of 20 minutes of involvement  per 1-minute annotated excerpt.
8 recordings were removed from the public distribution of the corpus in regard to ethical considerations (see section \ref{sec:criminal}), resulting in 277 1-minute long publicly available recordings.

The proposed corpus annotation protocol, operating on 1 minute out-of-context audio excerpts, 
resulted in several annotators going back and forth between annotation software and upload platform, and relatively high annotation times compared to other annotation campaigns: \cite{bazillon2008manual} reporting 8 hours of manual transcription and speaker turn segmentation per hour of annotated program.
This choice was motivated by concerns about obtaining a minimal amount of annotation errors.
While the display of an external automatic output may bias resulting annotations (annotators will only correct the portion of the signal where their production differs from the automatic system), we considered this approach offers less annotation biases than semi-automatic approaches  \cite{rats,salmon2014effortless,voxceleb2,uro2022semi} that may result in advantaging the system used for the pre-annotations, such as keeping the approximate time-code predicted by machines or discarding atypical speech events.



\section{Corpus Analysis}
\label{sec:corpusanalysis}

\subsection{Global Description}

\begin{table}
\centering
\begin{tabular}{lll}
\hline
\textbf{category} & \textbf{duration} & \textbf{speech \%} \\
\hline
generalist radio & 49 & 88.6 \\
music radio & 61 & 16.9\\
generalist TV     & 110 & 59.2 \\
news TV & 60 & 82.5 \\
\hline
\end{tabular}
\caption{\label{tab:globaldesc} Annotated data duration (minutes) and speech percentage per channel categories}
\end{table}

\begin{table}
\centering
\begin{tabular}{llllll}
\hline
\textbf{Corpus} & \textbf{dur} & \textbf{sp\%} & \textbf{mssd} & \textbf{mnsd}\\
\hline
AVA-Speech & 40h & 52.5\% & 1.76 & 1.48 \\
DIHARD2 & 47h33 & 75.7\% & 1.13 & 0.48 \\
REPERE & 39h22 & 93.5\% & 5.3 & 0.49 \\
ESTER 1+2 & 173h & 97.3\% & 8.33 & 1.99 \\
\textbf{InaGVAD} & 4h37 & 60.1\% &  2.22 & 0.59 \\
\hline
\end{tabular}
\caption{\label{tab:corpuscomp} Speech corpora description based on annotated duration (dur), speech percentage (sp\%), median speech segment duration (mssd) and median non-speech segment duration (mnsd)}
\end{table}

A raw description of \texttt{inaGVAD} annotated corpus is provided in table \ref{tab:globaldesc}, highlighting major differences between channel categories based on their annotated speech percentage, varying between 16.9\% in music radios and 88.6\% in generalist radios.

Table \ref{tab:corpuscomp} provides a summary of InaGVAD statistics in comparison to several speech resources described in section \ref{sec:related}.
Corpora designed for ASR (REPERE~\citeplanguageresource{giraudel2012repere} and ESTER~\citeplanguageresource{galliano2009ester}) have much highest annotated speech ratios (93.5 and 97.3\%) and longest median speech segment duration (5.3 and 8.33 seconds).
This highlights audiovisual ASR corpora tendencies to discard portions of the signal associated to low lexical density and focus on lexical content accuracy rather than on short non-speech events timing.
Median speech and non-speech segment duration found in InaGVAD (2.22 and 0.59 seconds) is comparable to those found in AVA-Speech~\citeplanguageresource{Chaudhuri2018} and DIHARD2~\citeplanguageresource{ryant2019second} VAD-oriented corpora.
Small duration differences observed may be partly explained by the different nature of these annotations.
DIHARD 2 is designed for diarization and splits speech segments corresponding to different speakers at speech turn boundaries, AVA-Speech splits speech segments depending on overlapped non-speech events, while InaGVAD considers adjacent speech turns uttered by speakers with the same characteristics (\textit{i.e.} 2 adult females) as a single speech segment. 
Conversely, InaGVAD categorization of non-speech events may explain its small non-speech segment duration. 
InaGVAD duration statistics being consistent with VAD standards, combined with high variability of vocal styles and background noises, makes it a very valuable and challenging resource for audiovisual VAD and SGS research.

\subsubsection{Speech Label Distribution}

\begin{table*}[hbt!]
\centering
\setlength{\tabcolsep}{5pt}
\begin{tabular}{|l|c|cccc|ccc|ccc|}
\hline
\textbf{channel} & \textbf{speech}& \multicolumn{4}{c|}{\textbf{speaker gender}} & \multicolumn{3}{c|}{\textbf{speaker age}} & \multicolumn{3}{c|}{\textbf{speech quality}} \\
\textbf{category} & \textbf{overlap} & \textbf{female} & \textbf{male} & \textbf{IDK} & \textbf{M+F} & \textbf{child} & \textbf{adult} & \textbf{senior} & \textbf{\{\}} & \textbf{-} & \textbf{*}\\
\hline
generalist radio & 3.2 & 21.8 & 77.3 & 0.0 & 0.9 & 0.3 & 95.0 & 4.5 & 97.5 & 1.0 & 1.4 \\
generalist tv & 6.3 & 38.0 & 58.2 & 2.6 & 1.1 & 2.7 & 92.7 & 2.7 & 96.6 & 1.8 & 1.5 \\
music radio & 2.2 & 44.2 & 53.8 & 0.0 & 2.0 & 0.0 & 100.0 & 0.0 & 97.8 & 0.3 & 0.0 \\
news tv & 8.9 & 38.7 & 57.7 & 0.0 & 3.6 & 0.0 & 97.3 & 2.1 & 98.6 & 1.4 & 0.0 \\
\hline
\end{tabular}
\caption{\label{tab:speechdetail} Speech labels distribution (duration \%) across channel categories (labels detailed in table \ref{tab:speech})}
\end{table*}

\begin{table*}
\centering
\begin{tabular}{|lccccccccccc|}
\hline
\textbf{category} & \textbf{overlap} & \textbf{AP} & \textbf{BR} & \textbf{BH} & \textbf{JI} & \textbf{MU1} & \textbf{MU2}  & \textbf{RE}  & \textbf{RI}  & \textbf{AU}  & \textbf{\{\}}\\
\hline
generalist radio & 1.0 & 0.6 & 11.3 & 0.0 & 5.6 & 14.6 & 5.0 & 38.5 & 2.5 & 0.0 & 22.9 \\
generalist tv & 7.8 & 1.3 & 25.1 & 3.2 & 0.6 & 26.5 & 29.0 & 4.5 & 2.6 & 1.5 & 14.0 \\
music radio & 0.0 & 0.0 & 0.3 & 0.1 & 1.4 & 95.0 & 1.9 & 0.8 & 0.0 & 0.0 & 0.5 \\
news tv & 3.3 & 0.0 & 15.9 & 0.6 & 5.4 & 16.3 & 23.3 & 20.5 & 0.0 & 0.1 & 21.3 \\
\hline
\end{tabular}
\caption{\label{tab:nonspeechdetail} Non-speech labels distribution (duration \%) across channel categories (labels detail in table \ref{tab:nonspeech})}
\end{table*}

A detailed analysis of speech label distribution is presented in table \ref{tab:speechdetail}.
Highest amounts of speech overlaps were found in news TV (8.9\%) while lowest rations were found  in music radio (2.2\%).
Speech overlaps involving speakers with different genders are reported using the \texttt{M+F} gender category, while overlaps involving same gender speakers are reported in their corresponding gender category.
In accordance with several studies~\cite{doukhan2018describing}, female speakers were found to be less represented than male speakers for each category of channels, with lowest speech-time ratio observed in generalist radios (21.8\%) and highest in music radio (44\%) - musical radio being the channels containing the lowest amount of speech.
\texttt{IDK} gender labels (associated to situations when annotators were not able to distinguish speaker gender)  were only observed in generalist TV channels (most probably related to cartoons) with a low frequency (2.6\%).
Speakers found in the corpus were mostly perceived as adults (92.7-100\%), with the biggest amount of child voices found in generalist TV (2.7\%) and highest amount of elderly voices in generalist radios (4.5\%).
Biggest amounts of atypical voice quality were found in generalist radio and generalist tv (1.4 and 1.5\%).

\subsection{Non-speech Label Distribution}

Table \ref{tab:nonspeechdetail} presents the distributions of annotated non-speech events. 
Generalist TV exhibits the most varied distribution of non-speech events, representing all categories defined in our annotation scheme, with the highest rate of non-speech-event overlap (7.8\%).
It features the highest proportion of noises (25.1\%), high amounts of foreground and background music (26.5\% and 29\%) and a reasonable amount of empty non speech events (14\%).
Musical radio non-speech events were unsurprisingly associated with large amounts of foreground music (95.0\%).
Conversely, generalist radio and news TV, with the lowest amount of non-speech events, mainly consist of unannotated non-speech events (22.9 and 21.3\%), audible respiration (38.5 and 20.5\%), noise (11.3 and 15.9\%) and jingles (5.6 and 5.4\%).






\section{Baseline System Benchmark}
\label{sec:eval}
\subsection{Dev and Test Set Split}
\label{subsec:devtest}

The corpus is split into two mutually-exclusive subsets to allow system comparison in fair and reproducible conditions. 
Development subset (\textbf{dev}) was obtained by randomly selecting 15 minutes for each of the 4 channel categories, resulting in 1h of annotated data that can be used to train, fine-tune, or adjust system parameters.
Testing subset (\textbf{test}) contains the remaining recordings (3h37) and should be used for obtaining evaluation results in the same conditions as the results presented in the following subsections.

\subsection{Voice Activity Detection}

\begin{table*}[hbt!]
\centering
\small
\setlength{\tabcolsep}{4.5pt}
\begin{tabular}{l|rrr|rrr|rrr|rrr}
\hline
 & \multicolumn{3}{c|}{\textbf{generalist radio}} & \multicolumn{3}{c|}{\textbf{music radio}} & \multicolumn{3}{c|}{\textbf{generalist tv}}  & \multicolumn{3}{c}{\textbf{news tv}}\\
\textbf{Open-source VAD} & \textbf{Acc} & \textbf{Prec} & \textbf{Rec} & \textbf{Acc} & \textbf{Prec} & \textbf{Rec} & \textbf{Acc} & \textbf{Prec} & \textbf{Rec} & \textbf{Acc} & \textbf{Prec} & \textbf{Rec}\\
\hline
\texttt{InaSpeechSegmenter} & 95.4 & 95.5 & \textbf{99.8} & \textbf{98.1} & \textbf{88.4} & 97.8 & 88.3 & \textbf{87.7 }& 93.6 & 95.0 & 95.1 & 99.5 \\
\texttt{LIUM\_SpkDiarization} & 94.3 & 94.8 & 99.4 & 74.2 & 32.2 & 98.0 & 81.4 & 80.6 & 90.7 & 92.1 & 92.7 & 98.8 \\
\texttt{Pyannote} &\textbf{ 96.3} & 96.7 & 99.4 & 75.7 & 33.8 & \textbf{99.9} & \textbf{89.5} & 86.5 & \textbf{97.7} & \textbf{96.1} & 96.1 & 99.6 \\
\texttt{Rvad} & 95.1 & \textbf{97.9} & 96.8 & 18.8 & 12.9 & 96.6 & 77.5 & 74.4 & 95.1 & 91.8 & 94.6 & 96.2 \\
\texttt{Silero} & 96.1 & 97.3 & 98.6 & 61.6 & 23.9 & 95.7 & 86.9 & 83.9 & 96.7 & 95.7 & \textbf{97.0} & 98.3 \\
\texttt{SpeechBrain} & 94.1 & 94.8 & 99.1 & 92.9 & 64.2 & 96.6 & 87.1 & 84.8 & 95.6 & 94.2 & 94.0 & \textbf{99.7} \\ 
\hline
\end{tabular}
\caption{\label{tab:vaddetail} Voice Activity Detection Accuracy, Precision and Recall on inaGVAD test set}
\end{table*}

\begin{table*}[hbt!]
\centering
\small
\begin{tabular}{lrrrrrrrrrr}
\hline
\textbf{Open-source VAD} & \textbf{AP} & \textbf{BR} & \textbf{BH} & \textbf{JI} & \textbf{MU1} & \textbf{MU2} & \textbf{RE} & \textbf{RI} & \textbf{AU} & \textbf{\{\}} \\
\hline
\texttt{InaSpeechSegmenter} & 15.4 & 35.6 & 36.0 & 8.1 & \textbf{1.5} & 25.9 & 95.1 & 77.0 & \textbf{22.0 }& 45.7 \\
\texttt{LIUM\_SpkDiarization} & 28.0 & 34.0 & 35.2 & 26.2 & 31.1 & 32.7 & 96.0 & 65.9 & 22.2 & 66.5 \\
\texttt{Pyannote} & \textbf{7.5} & \textbf{21.9} & 30.8 & \textbf{5.0} & 27.7 & \textbf{21.2} & 89.0 & \textbf{63.1} & 35.9 & 41.5 \\
\texttt{Rvad} & 33.4 & 33.8 & 60.8 & 83.8 & 91.5 & 53.8 & \textbf{41.1} & 85.7 & 68.3 & \textbf{21.7} \\
\texttt{Silero} & 10.7 & 25.9 & \textbf{29.8} & 20.1 & 41.1 & 25.7 & 63.4 & 76.0 & 33.4 & 35.4 \\
\texttt{SpeechBrain} & 30.3 & 35.1 & 40.0 & 11.9 & 9.3 & 29.0 & 97.7 & 72.7 & 21.2 & 66.0 \\
\hline
\end{tabular}
\caption{\label{tab:vadfa} Speech False Alarm rate per category of non-speech events (categories detailed in table \ref{tab:nonspeech}) }
\end{table*}

\label{sec:vadmetric}
VAD performances are described with pyannote metrics~\cite{pyannote.metrics}: Accuracy (proportion of input signal correctly classified), Precision (percentage of annotated speech in speech predictions), Recall (percentage of annotated speech detected), and False Alarm (non-speech incorrectly classified as speech).
A \textit{collar} value of 300 ms is used to remove from evaluation zones around reference annotation boundaries (150ms before and after), which seems reasonable with respect to the inaGVAD annotation process.


Six freely available VAD software, used with default parameters, were evaluated on inaGVAD test set: 
\texttt{LIUM\_SpkDiarization}~\cite{rouvier2013open}, \texttt{inaSpeechSegmenter}~\cite{doukhan2018ina}, 
\texttt{pyannote.audio}'s  VAD~\cite{Bredin2021},
\texttt{silero-vad}~\cite{SileroVAD},
\texttt{rvad} slow implementation~\cite{tan2020rvad}
and \texttt{SpeechBrain} \cite{speechbrain}.

\begin{table}[hbt!]
\centering
\begin{tabular}{lrrr}
\hline
\textbf{Open-source VAD} & \textbf{Acc} & \textbf{Prec} & \textbf{Rec} \\
\hline
\texttt{InaSpeechSegmenter} & \textbf{93.0} & \textbf{91.7} & 97.0 \\
\texttt{LIUM\_SpkDiarization} & 83.9 & 81.0 & 95.5 \\
\texttt{Pyannote} & 88.8 & 84.9 & \textbf{98.8} \\
\texttt{Rvad} & 69.8 & 67.4 & 95.9 \\
\texttt{Silero} & 84.4 & 80.5 & 97.6 \\
\texttt{SpeechBrain} & 90.9 & 88.4 & 97.7 \\
\hline
\end{tabular}
\caption{\label{tab:vadglobal} Voice Activity Detection Accuracy, Precision and Recall on inaGVAD test subset}
\end{table}



Table \ref{tab:vadglobal} presents global VAD results obtained with baseline systems.
\texttt{inaSpeechSegmenter} shows superior Accuracy and Precision (93 and 91.7) than other systems, while \texttt{pyannote} obtained the best recall (98.8).
Channel category results presented in table \ref{tab:vaddetail} provide more relevant insights into VAD challenges associated with audiovisual materials of different natures.
It is worth noting that no VAD system has succeeded in achieving the best results across all metrics for a given channel category.
As expected, better overall results were obtained for generalist radio and news TV categories (mean accuracy = 95.2 and 94.2), which are the most widely represented materials in annotated speech resources.
Largest differences between systems were observed for music radio (accuracy varying between 18.8 and 98.1).
Regardless of the system considered, generalist TV was shown to be the most challenging channel category, with best accuracy and recall obtained with pyannote (89.5 and 97.7) and best precision obtained with inaSpeechSegmenter (87.7).
These characteristics make it a particularly valuable material for evaluating audiovisual VAD systems.

In-depth speech false alarm analysis per non-speech event categories is presented in table \ref{tab:vadfa}. 
Relatively low represented in the corpus, laughs were found to be the most challenging non-speech event category, with best results obtained with pyannote (FA = 63.1).
Respiration was very challenging for several systems, most probably because these very short events are difficult to detect with large analysis windows, best results being obtained with \texttt{Rvad} (FA=41.1).
Largest differences between systems are observed on foreground music, with worst results obtained with rvad (FA=91.5) and best results obtained with inaSpeechSegmenter (FA=1.5).

The huge VAD performance disparity observed on music radios can be explained by several factors : 
The distinction between singing voice (annotated as non speech event) and spoken voice can be particularly challenging depending on musical genre~\cite{doukhan2017investigating}.
Moreover, VAD systems based on signal periodicity features 
(associating periodicity with speech and aperiodicity with noise) are not relevant for very periodic music material.
Lastly, some VAD software may use different definitions of speech including singing voice.
With regards to applications aimed at estimating women's speaking time percentage in media, we considered singing voice as a phenomenon that should be monitored separately from spoken voice. Moreover, it may require dedicated systems since gender prediction from singing voice is known to be harder than from spoken voice~\cite{kong2023strada}.

\begin{table*}[hbt!]
\center
\small
\begin{tabular}{l|ccc|ccccc|ccccc}
\hline
\textbf{SGS} & \multicolumn{3}{c|}{\textbf{Global Metrics}} & \multicolumn{5}{c|}{\textbf{Female metrics}} & \multicolumn{5}{c}{\textbf{Male metrics}}\\
\textbf{model} & \textbf{IER} & \textbf{Wrms} & \textbf{Werr} & \textbf{IER} & \textbf{rec} & \textbf{fa} & \textbf{md} & \textbf{conf} & \textbf{IER} & \textbf{rec} & \textbf{fa} & \textbf{md} & \textbf{conf} \\
\hline
\textbf{InaSS} & 14.6 & 13.3 & \textbf{-0.1} & 15.1 & 94.1 & 7.4 & 3.0 & 4.8 & 14.3 & 94.7 & 10.1 & 2.6 & 1.6 \\
\textbf{LiumSpk} & 36.8 & 29.1 & -6.0 & 56.6 & 88.2 & 32.4 & 4.7 & 19.5 & 25.9 & 85.7 & 18.4 & 3.6 & 3.9 \\
\textbf{ECASGS} & \textbf{14.0 }& \textbf{11.0} & 1.8 & 10.7 & 92.6 & 6.2 & 1.1 & 3.4 & 15.8 & 97.0 & 11.1 & 1.2 & 3.5 \\
\hline
\end{tabular}
\caption{\label{tab:genderdetail} SGS results detailed per gender obtained with \texttt{InaSpeechSegmenter} (InaSS), \texttt{LIUM\_SpkDiarization} (LiumSpk) and \texttt{ECASGS} (ours)}

\end{table*}

\begin{table*}[hbt!]
\center
\small
\begin{tabular}{l|ccc|ccc|ccc|ccc}
\hline
  \textbf{SGS} & \multicolumn{3}{c|}{\textbf{generalist radio}} & \multicolumn{3}{c|}{\textbf{music radio}} & \multicolumn{3}{c|}{\textbf{generalist tv}} & \multicolumn{3}{c}{\textbf{news tv}}\\

 \textbf{model} & \textbf{IER} & \textbf{Werr} & \textbf{Wrms} & \textbf{IER} & \textbf{Werr} & \textbf{Wrms} & \textbf{IER} & \textbf{Werr} & \textbf{Wrms} & \textbf{IER} & \textbf{Werr} & \textbf{Wrms}\\ \hline
\textbf{InaSS} & \textbf{5.9} & \textbf{0.7} & \textbf{4.4} & \textbf{16.3} & 3.6 & 15.8 & 25.5 & \textbf{-1.4} & 16.6 & \textbf{6.5} & \textbf{1.0} & \textbf{4.3} \\
\textbf{LiumSpk} & 9.8 & -2.4 & 11.3 & 219.1 & -3.6 & 36.8 & 47.3 & -10.1 & 32.6 & 15.9 & -2.1 & 20.7 \\
\textbf{ECASGS} & \textbf{5.9} & 1.6 & 6.8 & 16.7 & \textbf{2.9} & \textbf{12.2} & \textbf{23.9} & 2.5 & \textbf{13.4} & \textbf{6.5} & 1.7 & 5.0 \\
\hline
\end{tabular}
\caption{\label{tab:gendercat} SGS results accross channel categories obtained with \texttt{InaSpeechSegmenter} (InaSS), \texttt{LIUM\_SpkDiarization} (LiumSpk) and \texttt{ECASGS} (ours)}
\end{table*}

\subsection{Speaker Gender Segmentation}
\label{sec:sgs}

\subsubsection{SGS Evaluation Metrics}

SGS is evaluated using pyannote Identification Error Rate (\textbf{IER})~\cite{pyannote.metrics}  :
$$IER = \frac{fa + md + conf}{speech~duration}$$
with \textbf{fa}, \textbf{md} and \textbf{conf} corresponding to False Alarm, Missed Detection and Confusions.
IER extensions are proposed to detail performances for each gender such as Male confusion refers to Male speech incorrectly predicted as Female speech, Female false alarm to non-speech incorrectly predicted as Female speech,  and Female missed detection to Female speech predicted as non-speech.
IER-related estimates ignore portions of signal where speaker gender cannot be inferred from annotations (IDK gender label or overlapped male and female voices) and use a collar value of 300ms (see section \ref{sec:vadmetric}).

Additional metrics describing Women Speaking Time Percentage \textbf{WSTP} 
estimation robustness  are reported ($WSTP=40\%$ corresponding to $40\%$ of women speech and $60\%$ of male speech - excluding non-speech events).
\textbf{Werr} refers to WSTP global estimation error on a given subset and is defined as the difference between reference and predicted WSTP : positive Werr means underestimated women speaking-time percentage while negative Werr means overestimated WSTP.
\textbf{Wrms} refers to the WSTP root mean square error per recording : lowest Wrms are better and Wrms is generally lower for long recordings~\cite{doukhan2018open}.
Speech segments without annotated gender information (IDK label : max: 3.6\% in generalist TV) were taken into account in WSTP metrics, adding half of their duration to reference male and other half to reference female speech time.
In the absence of annotated or predicted speech in a recording, WSTP was set to 50\%.

\subsubsection{SGS Baseline Models}

Two freely available SGS systems were evaluated on InaGVAD test set using their own VAD module : \texttt{InaSpeechSegmenter}~\cite{doukhan2018open} and \texttt{LIUM\_SpkDiarization}~\cite{rouvier2013open}.


Additionally, we proposed \textbf{ECASGS}, a baseline SGS trained on InaGVAD dev set.
This system is based on ECAPA-TDNN speaker embeddings~\cite{Desplanques-2020} obtained using a model pretrained on Voxceleb dataset~\citeplanguageresource{voxceleb1,voxceleb2} for speaker recognition\footnote{\url{https://huggingface.co/speechbrain/spkrec-ecapa-voxceleb}}.
Speaker embeddings were extracted from inaGVAD dev set using a sliding window of size of 3 seconds with 100 milliseconds shift, and used to train a Support Vector Machine with Radial basis function kernel for the binary male/female classification task.
This proposal depends on external VAD annotations and we use the best reported VAD for each channel category according to the accuracy performance metric (see table \ref{tab:vaddetail}): inaSpeechSegmenter VAD for music radio and pyannote VAD for remaining channel categories.


\subsubsection{SGS Results}

Table \ref{tab:genderdetail} presents global and detailed evaluation metrics obtained on inaGVAD test set.
InaSpeechSegmenter achieved a close-to-perfect Werr (-0.1) showing better abilities for large grain audiovisual WSTP analysis, despite False Alarm rates slightly higher for male than for female (+2.7) and larger female confusion (+3.2).
While ECASGS managed to obtain the best performances according to the IER and Wrms criterions, it was associated with a Werr value of 1.8, showing a tendency to slightly underestimate women speaking time percentage. Larger Male False alarm rates (+4.9) and lower Female recall (-4.4) may be accounted for this tendency.
LiumSpkDiarization obtained the lowest performances, with a significant tendency to overestimate female voices (Werr=-6), which may be partly explained by a strong Female False Alarm rate (32.4\%).

Global results obtained per channel categories are detailed in table \ref{tab:gendercat}.
Due to their low amount of speech and important amount of gender predictions obtained from non-speech segments, music radios were associated with the highest Werr showing a tendency of  ECASGS (2.9) and inaSpeechSegmenter (3.6) to underestimate significantly WSTP.
While generalist radio and news TV present a relatively low SGS difficulty, Generalist TV was the most challenging material according to the IER criterion (best IER=23.9 with ECASGS).

\section{Conclusion}
\label{sec:ccl}

We presented inaGVAD, a diverse and freely-available audiovisual corpus annotated with extended VAD and speaker trait annotations.
Benchmarks realized across channel and non-speech event categories allowed to distinguish the different abilities of 6 state-of-the-art VAD systems, and highlighted the challenges associated with the design of automatic tools suited to the diversity of audiovisual programs: generalist TV being the most challenging channel category.
All VAD systems obtained lower VAD performances estimates on inaGVAD generalist tv and music radio subsets compared to the estimates reported by~\citealp{Bredin2021} on AMI, DIHARD 3 and VoxConverse VAD corpora.
This suggests that inaGVAD corpus  contains original materials compared to other state-of-the-art corpora (making it challenging and valuable), or at the very least, it may employ different speech/non speech definitions tailored to use cases that have not traditionally been addressed.
More in-depth analysis of gender predictions errors based on speaker traits (including speakers with IDK gender label) or speech quality goes beyond the scope of this paper but is addressed with dedicated evaluation scripts (see section \ref{sec:avail}).\\

We also show how a simple transfer-learning speaker gender segmentation baseline trained on inaGVAD one-hour development set could obtain competitive results on the test set, illustrating the relevance of our corpus for designing new VAD and SGS methods.
While SGS results on generalist radio and news TV are similar to those reported on REPERE corpus~\cite{doukhan2018open}, we show music radio and generalist TV subsets are more challenging materials, requiring additional research efforts, and we were pleased to note that our baseline model managed to improve Wrms for these challenging channel categories.\\

IaGVAD has a relatively low annotated data duration compared to other related corpora (see table \ref{tab:corpuscomp}). This limitation is partly due to budgetary considerations associated with a resource intended for a free-of-charge distribution (96 annotation hours not including quality control engineering and coordination time).
Moreover, inaGVAD annotation process (section \ref{sec:annotationprocedure}) included several back and forth between the annotation software and the upload and validation platform, resulting in notably higher annotation time (20 annotation minutes per 1-minute excerpt) compared to other studies~\cite{bazillon2008manual}.
Annotators informal feedback on this process described machines making prediction errors half the time, while they were doing annotation errors the other half, suggesting this choice improved the overall quality of annotations.
Our proposal favored audiovisual material variability, annotation scheme originality and annotation granularity rather than total duration.
The significance of this proposal is illustrated by the relatively low performances of state-of-the-art VAD and SGS systems on generalist TV and music radio subsets despite being trained and evaluated on large datasets.
Recent self-supervised learning advancements suggest a reduced dependency on extensive training data and this philosophy underscores our decision to divide inaGVAD into a development and test set, omitting a dedicated training set: the development dataset being mainly targeted for fine-tuning and parameter adjustments.
We assert that our corpus is best used with external training data sources and demonstrated this with a competitive baseline model (see section \ref{sec:sgs}).

Among the known limitations of our annotation scheme, annotators were not asked to report non-binary speaker gender identities and were provided three gender label options: male, female, and "I Don't Know".
IDK label was intended for cases where annotators were not able to assign speaker voices to binary gender categories, which is different from suspecting a non-binary gender identity.
It's worth noting that annotators reported they did not encounter voices exhibiting discernible acoustic correlates indicative of non-binary gender expression during the annotation campaign, which is consistent with the known limited representation of openly non-binary individuals in French TV and radio programs.
Additionally, while some speaking styles may convey elements of gender identity, these cues are contextually bound \cite{holmes1998signalling,kachel2024speakers}
and reliable annotation of gender identity requires explicit speaker statements. Our annotation campaign, based on out-of-context 1-minute long audio recordings, makes it difficult  to annotate non-binary gender identity based on purely acoustic cues.
While some studies reported non-binary continuous gender estimates based on crowd judgments from acoustic cues~\cite{doukhan2023voice,chen2020objective}, extensions of our  gender annotation scheme in single annotator configurations would require additional research to define reproducible speaker gender categories or continuous scales transcending the binary gender paradigm and associated with reasonably high inter-annotator agreements and annotation times.





\section{Corpus Availability and Research Reproducibility}
\label{sec:avail}
The research efforts presented in this study are aimed at being shared among the speech science research community.

Corpus sound files can be freely downloaded from INA's FTP server.
Users need to fill a request form and to accept general conditions of use (strictly restricting corpus access to academic research) accessible at 
\url{https://www.ina.fr/institut-national-audiovisuel/research/dataset-project} .

Code, annotations, and automatic baseline outputs associated with this paper are freely available on github under MIT license accessible at 
\url{https://github.com/ina-foss/InaGVAD} .
This repository contains routines allowing to train on dev set and evaluate on test set the proposed ECAPA-TDNN baseline, which may be used as a basis for designing new systems.
It also contains command-line programs allowing the evaluation of external systems in order to obtain VAD and Gender estimates which could be directly compared to the results presented in this paper.
The proposed codebase will be actively maintained to suit users' requests.

\section{Ethical Considerations}
\label{sec:ethical}

\subsection{Annotators Working Conditions}

Corpus annotation campaign was realized by audiovisual professionals having a permanent contract (French CDI).
Annotators received a presentation on gender representation in media based on a review of academic studies and reports, illustrating how automation (in particular VAD and SGS) could be used to analyze massive amounts of data.
This presentation contributed to enhance their understanding of the project's objectives and provide context for the annotation work, which could be perceived as monotonous.
Annotators had the flexibility to distribute their 48-hour workload dedicated to the annotation campaign over 6 weeks.
Early stages of the annotation campaign aimed at producing more detailed annotations, including speech transcription and speaker diarization.
Annotators reported difficulties to transcribe proper nouns and distinguish speakers in out-of-context one-minute long recordings, suggesting transcription and diarization annotation may be more suited to longer excerpts providing speaker and topic presentation.
Early stages of the annotation campaign included several meetings with authors, where ambiguous cases were collectively exposed and discussed, resulting in annotation instruction updates based on annotators' feedback.

\subsection{Criminal Conviction Filtering}
\label{sec:criminal}
In the context of making the corpus publicly available, INA's legal department has insisted on the removal of audio excerpts referencing criminal convictions or accusations. This request is grounded in Article 10 of the European General Data Protection Regulation (GDPR)~\footnote{\url{https://gdpr-info.eu/}}, which restricts the processing of personal data related to criminal convictions, offenses, or related security measures to official authorities and specific use cases. The primary concern is that the corpus may be requalified as a conviction database and declared unlawful. Given the absence of exemptions for processing criminal data for research purposes, even under Article 89 of the GDPR, a manual review was imperative to ensure the absence of any criminal data within the corpus.

A semi-automatic procedure, based on Whisper general-purpose speech recognition model \cite{radford2023robust}, was proposed to comply with this request and speed-up corpus manual inspection. 
Authors, with background in automatic speech analysis, were in charge of determining if recordings contain references to criminal convictions based on their automatic speech transcripts (about 33800 words). They were also in charge of deciding if some recordings should be listened to manually, if they suspect speech transcription errors, hallucinations, but also to help to distinguish between fictive (movies, series) and real criminal conviction.

The use of this semi-automatic process provided a valuable speed-up allowing to check 285 minutes of recordings in 133 minutes. By the end of this process, 8 recordings were discarded, mostly related to news programs or documentaries.

\section{Acknowledgements}

This work has been partially funded by \anonymize{the French National Research Agency (project Gender Equality Monitor - ANR-19-CE38-0012)}. The authors would also like to thank Jean-Hugues Chenot (INA) for conducting the repetition analysis, which helped avoid duplicate audio patterns in the development and test sets.

\section{Bibliographical References}\label{sec:reference}

\bibliographystyle{lrec-coling2024-natbib}
\bibliography{lrec-coling2024-example}

\section{Language Resource References}
\label{lr:ref}
\bibliographystylelanguageresource{lrec-coling2024-natbib}
\bibliographylanguageresource{languageresource}

\end{document}